\documentclass[conference]{IEEEtran}

\usepackage{blindtext, graphicx}
\usepackage{moreverb}
\usepackage{hyperref}
\usepackage{graphicx}
\usepackage{amssymb}
\usepackage{amsfonts}
\usepackage{multicol}
\usepackage{amsmath}
\usepackage{mathtools}
\usepackage[mathscr]{euscript}
\usepackage{booktabs}
\usepackage{makecell}
\usepackage{multirow}
\usepackage{amsbsy} 
\usepackage[hypcap=true, font=small]{caption}
\usepackage[lined,vlined,ruled]{algorithm2e}
\usepackage{cite}
\usepackage{xfrac}
\usepackage{bbm}
\usepackage{eufrak}
\usepackage{color, colortbl}
\definecolor{Gray}{gray}{0.9}
\usepackage[first=0,last=9]{lcg}

\long\def\/*#1*/{}

\usepackage{tikz}
\newcommand\copyrighttext{%
  \footnotesize \textcopyright~\the\year~IEEE. Personal use of this material is permitted.
  Permission from IEEE must be obtained for all other uses, in any current or future
  media, including reprinting/republishing this material for advertising or promotional
  purposes, creating new collective works, for resale or redistribution to servers or
  lists, or reuse of any copyrighted component of this work in other works.}
\newcommand\copyrightnotice{%
\begin{tikzpicture}[remember picture,overlay]
\node[anchor=north,yshift=-15pt] at (current page.north) {\fbox{\parbox{\dimexpr\textwidth-\fboxsep-\fboxrule\relax}{\copyrighttext}}};
\end{tikzpicture}
\vspace{-0.3cm}
}

\begin{document}

\title{On Reliability-Aware Server Consolidation in \\ Cloud Datacenters}

\author{\IEEEauthorblockN{Amir Varasteh$^\dagger$, Farzad Tashtarian$^\star$, and Maziar Goudarzi$^\S$ \\ \\
$^\dagger$ Chair of Communication Networks, Department of Electrical and Computer Engineering, \\
Technical University of Munich, Germany \\
\{Email: amir.varasteh@tum.de\}\\
$^\star$ Department of Computer Engineering, Mashhad Branch, Islamic Azad University, Mashhad, Iran, \\ \{Email: f.tashtarian@mshdiau.ac.ir\}}
$^\S$ Department of Computer Engineering, Sharif University of Technology, Tehran, Iran \\ \{Email: goudarzi@sharif.edu\}}

\maketitle

\thispagestyle{plain}
\pagestyle{plain}

\copyrightnotice{}

\begin{abstract}

In the past few years, datacenter (DC) energy consumption has become an important issue in technology world. Server consolidation using virtualization and virtual machine (VM) live migration allows cloud DCs to improve resource utilization and hence energy efficiency. In order to save energy, consolidation techniques try to turn off the idle servers, while because of workload fluctuations, these offline servers should be turned on to support the increased resource demands. These repeated on-off cycles could affect the hardware reliability and wear-and-tear of servers and as a result, increase the maintenance and replacement costs. In this paper we propose a holistic mathematical model for reliability-aware server consolidation with the objective of minimizing total DC costs including energy and reliability costs. In fact, we try to minimize the number of active PMs and racks, in a reliability-aware manner. We formulate the problem as a Mixed Integer Linear Programming (MILP) model which is in form of NP-complete. Finally, we evaluate the performance of our approach in different scenarios using extensive numerical MATLAB simulations.
\end{abstract}
\begin{IEEEkeywords}
Cloud computing, datacenter, energy optimization, cost optimization, vm placement, server consolidation, reliability.
\end{IEEEkeywords}

\IEEEpeerreviewmaketitle

\section{Introduction}

The role of Cloud computing and its applications in our daily life are growing exponentially. In this way, users can use these applications (e.g. search engines, email, file storage) without the need to own the service or infrastructure. These clouds provide wide range of services hosted by DCs in a "pay-as-you-go" manner, which helps organizations to reduce the CAPEX and OPEX costs and focus on their core business. 
Due to unpredictable and growing demand for Internet-based services and resources, DCs computing and storage capacities has been increased significantly. Consequently, there has been a rapid rise in energy consumption and carbon dioxide (CO$_2$) footprints of these DCs, which are a major challenge in both industry and academia \cite{beloglazov2011taxonomy}. \par
Physical resources, along with networking and cooling devices are the main power consumers in DCs. However, the average utilization of physical resources in cloud DCs is relatively low and it is between 10\% and 50\% \cite{hozle2007}. This could lead to massive energy wastage, because an idle server consumes at about 70\% of its peak draw \cite{chen2008energy,gandhi2009optimal}. To cope with this challenge, server consolidation technique is widely used in cloud DCs. This technique, which is working based on virtualization technology, pack DC virtual machines (VMs) on minimum number of Physical Machines (PMs) to improve resource utilization and decrease the energy consumption by shutting down idle servers \cite{padala2007performance}. 

On the other hand, because of over-aggressive consolidation methods, combined with DC workload fluctuations, the turned off servers would be turned on to serve the incoming workload. These repeated on-off cycles have several negative impacts on servers wear-and-tear and reliability (i.e. aging), and hence replacement and procurement costs: 1) Repeated high transition frequency and on-off cycles are  recognized as the most crucial factor impairing disk reliability \cite{xie2008sacrificing,xie2011understanding}. 2) On-off thermal cycle of CPU, which is another factor causing server failures \cite{srinivasan2005lifetime,srinivasan2010characterizing}. Therefore, repeated on-off cycles of PMs in consolidation approaches, increase the wear-and-tear of server components, incurring replacement and procurement costs and also partial or complete service(s) downtime that costs \$5,000 per minute \cite{bodik2012surviving}. Thus, in addition to \textit{short-term} energy savings, \textit{long-term} reliability and maintenance costs are also an important issue that needs to address. Hence, in this paper, the key question to answer is: \\
\textit{"Considering DC energy consumption (PMs, cooling, and network), reliability, and migration costs, how server consolidation should be performed to minimize the total DC costs"}
\par

In response, this paper presents a mathematical model with the objective of minimizing total DC costs. We analyze and characterize the energy and reliability costs in a DC. Using these costs, we formulate the above problem as a Mixed Integer Linear Programming (MILP) mathematical model which is in form of NP-complete. Moreover, we simulate the proposed approach in MATLAB software and then evaluate the performance of the presented approach through extensive simulation experiments.
Therefore, the main points of this paper could be summarized as follows: 1) Providing a mathematical model for reliability-aware server consolidation in cloud DCs. 2) Taking disk and CPU reliability impacts on PMs into account to provide a reliability-friendly server consolidation approach. 3) Considering rack structure (including network and cooling devices) In addition to PMs and VM migration costs for energy-efficiency purposes. \par

The rest of this paper is organized as follows: we start by discussing related work in literature (section II). In sections III, we present the system model and formulations. Then, we describe the proposed mathematical model in section III. The performance of the presented approach is evaluated in section V, and finally, section VI concludes the paper along with some future directions.

\begin{table}[t]
    \centering
    \small
    \caption {Notations}
    \label{table1}
    \begin{tabular*}{1\linewidth}{cl}
    \Xhline{2\arrayrulewidth}
    Notation & \hspace{2.5cm} Description \\ \hline
    $\mathbb{V}$ & Set of VMs, $\mathbb{V}=\{v_1,v_2,...,v_{\mathbb{V}}\}$\\
    $\mathbb{P}$ & Set of PMs, $\mathbb{P}=\{p_1,p_2,...,p_{\mathbb{P}}\}$\\
    $\mathbb{R}$ & Set of Racks, $\mathbb{R}=\{r_1,r_2,...,r_{\mathbb{R}}\}$\\
    $t$ & Time-slot index \\
    $\tau$ & Time-slot duration \\
    $C^{ene} $ & Total server consolidation energy cost\\
    $C^{rel} $ & Total server consolidation reliability cost\\
    $G^{rel}$ & Total server consolidation reliability gain \\
    $c^{pm}_i$ & Energy cost of $p_i$\\
    $c^{rack}_i$ & Energy cost of $r_i$\\
    $c^{mig}$ & Total energy cost of VM migrations\\
    $S'$, $S$ & VM-to-PM mapping matrices for time-slot $t$ and $t+1$\\
    $c^{disk}_i$ & Disk reliability cost for $p_i$\\
    $c^{cpu}_i$ & CPU reliability cost for $p_i$\\
    $c^{ToR}$ & ToR switch energy consumption\\
    $c^{cooling}$ & Rack cooling device energy consumption\\
    $R^u_i$ & $i^{th}$ VM requirement for resource type $u \in U$\\
    $\bar{C}_i^u$ & $i^{th}$ PM total capacity of resource type $u \in U$ \\
    \Xhline{2\arrayrulewidth}
    \end{tabular*}
\end{table}

\section{Related Work}
The server consolidation technique determines the VM-to-PM mapping for a DC with the aim of minimizing the number of online PMs. In fact, it packs the VMs on minimum number of PMs to save the energy by turning off the idle PMs. However, this problem could be tackled by considering various parameters and/or objectives \cite{varasteh2015server}. In the following, we categorize these problem types and present some of the recent works in the literature. \par
\textit{Performance Awareness:} Current virtualization techniques do not guarantee efficient performance isolation between VMs hosting on a PM \cite{roytman2013pacman}. The contention in resources such as shared caches and memory bandwidth could lead to performance degradation and hence, Service Level Agreements (SLAs) violations \cite{tickoo2010modeling}. There are several works in literature that have considered the inter-VM performance degradation in design of their server consolidation algorithms \cite{nathuji2010q,roytman2013pacman}. They used decent performance profiling methods to compute performance degradation of any possible collocated VM combinations on a PM. Based on that, they allocated the VMs with less performance interference on a certain PM. In this way, in addition to energy consumption, performance interference would be also minimized. \par
\textit{Traffic Awareness:} Conventional server consolidation approaches have not considered the traffic/communications among VMs in the DC. This can lead to situations where heavy traffic transfers between pairs of VMs that are placed on PMs far from each other (e.g. different pods or racks) and impose large traffic cost to the DC \cite{meng2010improving}. To cope with this challenge, for instance, authors in \cite{meng2010improving} proposed an algorithm that use the traffic matrix among the VMs and the communication cost matrix among PMs as input. The algorithm then places the DC VMs on appropriate PMs with the aim of minimizing the traffic passing through DC network switches.\par
\textit{Reliability Awareness:} There are two general aspects of reliability in server consolidation approaches: 1) service reliability, and 2) hardware reliability costs \cite{varasteh2015server}, which in this paper, we focus on the second category. Authors in \cite{guenter2011managing}, used a Markov state model in order to satisfy the workload demands, while minimizing the energy and reliability costs due to repeated on-off cycles. Also, \cite{deng2014reliability} presented a reliability-aware server consolidation approach which used a grouping genetic algorithm (GGA) to minimize total DC operational costs. \par
However, to the best of our knowledge, this the first paper that provides a mathematical model for reliability-aware server consolidation with the aim of minimizing total DC costs, considering energy (including PMs, cooling, and network devices in racks), reliability costs, and migration costs all together.

\begin{figure}[t]
\includegraphics[width=\linewidth]{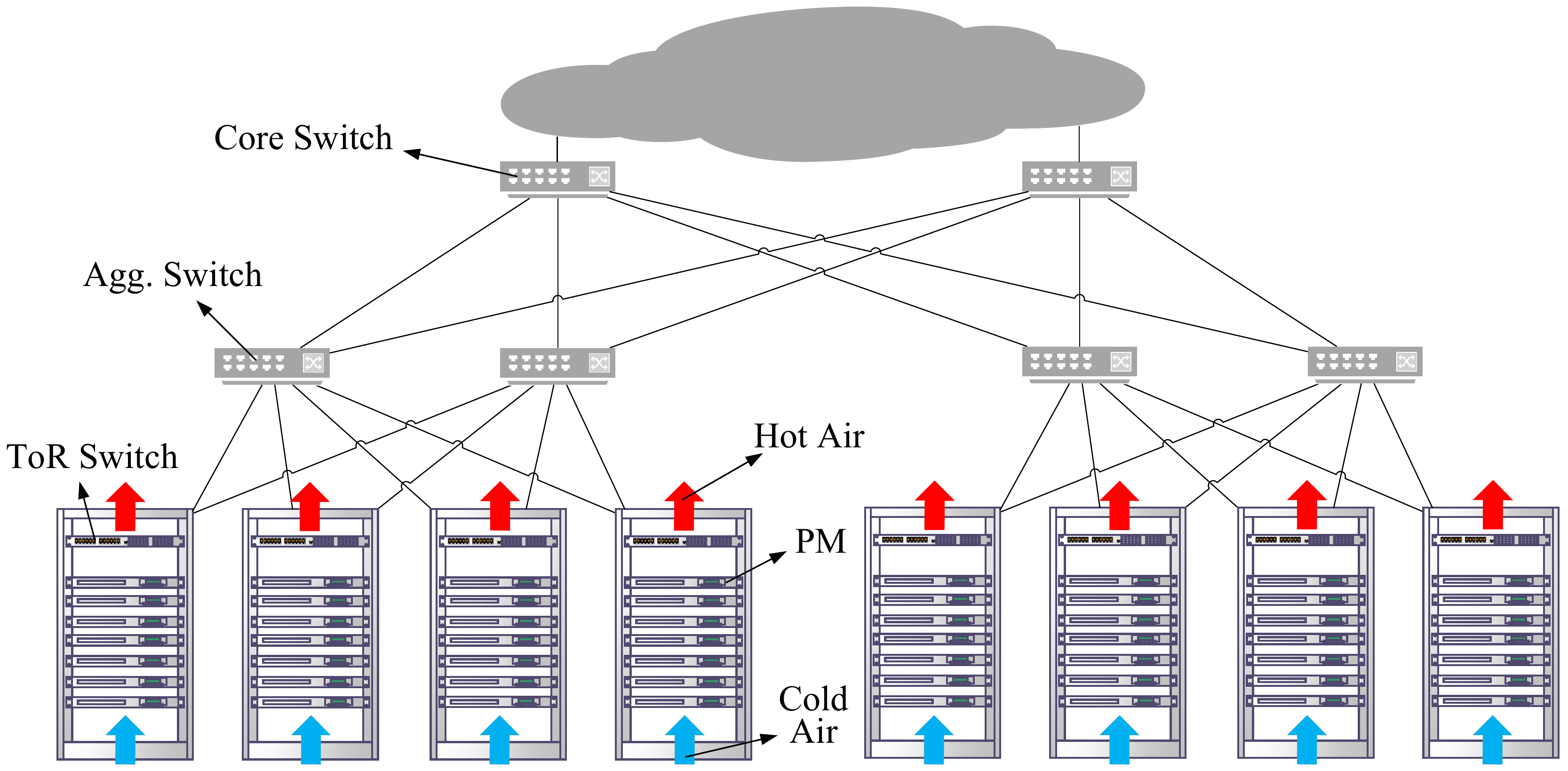}
\caption{A typical fat-tree datacenter with rack-based cooling system}
\label{fig1}
\end{figure}

\section{System Model}
We consider a DC with a typical fat-tree topology and a set of PMs which is defined as $\mathbb{P}= \{ p_1, p_2, ... , p_{|\mathbb{P}|} \}$. To simplify the problem, we consider the DC as homogeneous. The DC PMs are stored in DC racks defined as $\mathbb{R}= \{ r_1, r_2, ... , r_{|\mathbb{R}|} \}$. Each rack includes a rack-based cooling system, in which cold air is delivered directly inside the rack and the hot air exits the rack using heat risers. This cooling system increases the power efficiency since the air recirculation in conventional cooling systems is removed, and the air conditioner is brought closer to the target (i.e. PMs) \cite{esfandiarpoor2015structure}. There is also a top of rack (ToR) switch and a number of PMs in each rack (see Fig. \ref{fig1}). There are also a set of VMs in DC defined as $\mathbb{V}=\{ v_1,v_2,...,v_{|\mathbb{V}|}  \}$ which are hosted by DC PMs. Without loss of generality, we consider a discrete time model, in which the time is slotted into equal time slots denoted by $t$ with the duration of $\tau$. The server consolidation algorithm is run at the beginning of each time-slot. For clarifications, the main parameters that are used in mathematical formulations are described in Table \ref{table1}.

\section{Proposed Mathematical Model}

In this section, we present the proposed mathematical model for reliability-aware server consolidation problem. The following model will run by the DC \textit{resource management} framework in each time-slot $\tau$. In this server consolidation approach, considering the current time-slot $t$, using the proposed model, we determine the VM-to-PM mapping for DC in time-slot $t+1$. We consider three cost components to contribute in total DC cost. Let us first define the objective function of the proposed mathematical model:
\begin{align}
    &\textbf{Minimize} \hspace{1cm}  \alpha C^{ene} + \beta C^{rel} - \gamma G^{rel} \nonumber
\end{align}
where $C^{ene}$, $C^{rel}$, and $G^{rel}$ are the values for total DC energy cost, reliability cost, and reliability gain for the determined VM-to-PM mapping in time-slot $t+1$, respectively. Also, $\alpha$, $\beta$, and $\gamma$ are weighting factors to adjust the relative importance of the cost components, which are between 0 and 1.\par
\begin{figure}[t]
\includegraphics[width=\linewidth]{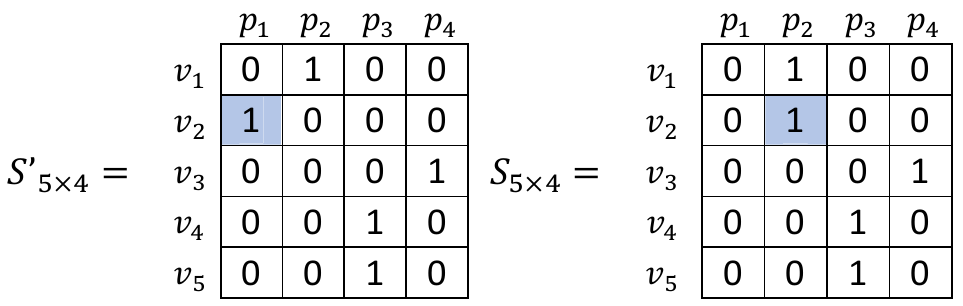}
\caption{$S'$ and $S$ matrices.}
\label{fig2}
\end{figure}

Before formulating these costs and modeling the related constraints, let us define general variables and constraints. Suppose $S'_{|\mathbb{V}| \times |\mathbb{P}|}$ and $S_{|\mathbb{V}| \times |\mathbb{P}|}$ be the matrices to show the DC VM-to-PM mapping for time-slot $t$ and $t+1$, respectively. Fig. \ref{fig2} shows an example of $S'$ and $S$ for 4 PMs and 5 VMs. For instance, $S'_{21}=1$ in Fig. \ref{fig2} states that $v_2$ is hosted by $p_1$ in time-slot $t$, and after running server consolidation algorithm, it is migrated to $p_2$ in time-slot $t+1$, and $p_1$ turns off (i.e. $\sum_{i=1}^{|\mathbb{V}|} \nolimits S_{ij} = 1$, and $j=1$). Moreover, let us define the  binary variables $F^{00}_i$ and $F^{10}_i$. $F_i^{00}=1$ if $p_i$ is offline in time-slot $t$ and remains offline in time-slot $t+1$ (i.e. $\sum_{j=1}^{|\mathbb{V}|} \nolimits S'_{ji} = 0$ and $\sum_{j=1}^{|\mathbb{V}|} \nolimits S_{ji} =0$), otherwise equals to $0$. Similarly, $F_i^{10}=1$ if $p_i$ is online in time-slot $t$ and powers off in time-slot $t+1$ (i.e. $\sum_{j=1}^{|\mathbb{V}|} \nolimits S'_{ji} > 0$ and $\sum_{j=1}^{|\mathbb{V}|} \nolimits S_{ji}=0$), otherwise equals to $0$. Obviously, these values can be easily obtained using $S$ and $S'$ matrices. In fact, there are four transition  states for any $p_i \in \mathbb{P}$ from $t$ to $t+1$: \textit{offline} to \textit{online}, \textit{offline} to \textit{offline}, \textit{online} to \textit{offline}, and \textit{online} to \textit{online}. However, we can model the targeted problem only by using the first two transition states. Additionally, suppose $p_i$ is \textit{offline} in time-slot $t$. Thus, the value of $F^{10}$ has to be equal to 0. On the other hand, if $p_i$ is \textit{online} in time-slot $t$, the value of $F^{00}$ must be equal to 0. Therefore, we define them as the following two constraints for the proposed model:
\begin{align}
    &F_i^{10} = 0, \text{ } \forall i \in \mathbb{P},\text{ if $p_i$ is offline in $t$} \label{eq6} \\
    &F_i^{00} = 0, \text{ } \forall i \in \mathbb{P}, \text{ if $p_i$ is online in $t$} \label{eq7}
\end{align}

Notably, the optimal values for $F_i^{10}$ and $F_i^{00}$ for online and offline PMs are determined by running the model. As the next constraints in the model, when the model does not determine any VMs on a certain PM at time-slot $t+1$, it means that the sum of $F^{10}_i$ and $F^{00}_i$ must be equal to 1. Therefore, to ensure the consistency between $F^{10}_i$, $F^{00}_i$ and $S$ matrix, the following constraints should be defined:
\begin{align}
    &S_{ij} \leq 1 - (F_j^{10} + F_j^{00}), \hspace{.3cm} \forall i \in \mathbb{V}, \forall j \in \mathbb{P} \label{eq8} \\
    &1-(F^{10}_j + F_j^{00}) \leq \sum_{i=1}^{|\mathbb{V}|} \nolimits S_{ij}, \text{ } \forall j \in \mathbb{P} \label{eq9}
\end{align}

Additionally, a capacity constraint has to be satisfied in the proposed mathematical model: the sum of VMs resource requirements should not be more than the total PM resource capacity which is defined as follows:
\begin{align}
    \sum_{i=1}^{|\mathbb{V}|} \nolimits S_{ij} R_i^u \leq \bar{C}_j^u, \hspace{.5cm} \forall j\in \mathbb{P}, \text{ } \forall u \in U \label{eq15}
\end{align}
where $R_i^u$ is the $i^{th}$ VM demand for resource type $r \in U$, where $U$ is the set of resource types, e.g. CPU, RAM. Also, each VM must be hosted by only one PM. This constraint can be wrote as follows:
\begin{align}
    \sum_{j=1}^{|\mathbb{P}|} \nolimits S_{ij}  =  1, \hspace{.5cm} \forall i \in \mathbb{V} \label{eq16}
\end{align}
Now, let us define the energy saving constraints for our model. As the first constraint, suppose $p_i$ is located in rack $r_k$. In this case, if there is not any VMs on a that PM, the PM should be turned off:
\begin{align}
    \sum_{i=1}^{|\mathbb{V}|} \nolimits S_{ij} \leq X_{jk}  \mathscr{M}, \hspace{.5cm} \forall j \in \mathbb{P}, \text{ } \forall k \in \mathbb{R} \label{eq17}
\end{align}
where the binary variable $X_{jk} = 0$ if the $j^{th}$ PM which is stored in $r_k$ must be turned off in time-slot $t+1$. Also, $\mathscr{M}$ is set to a large positive value. The next equation is the rack control constraint. This constraint is presented for any $p_j \in \mathbb{P}$, if it is located in $r_k$:
\begin{align}
    \sum_{j=1}^{|\mathbb{P}|} \nolimits X_{jk} \leq Y_k  \mathscr{M}, \hspace{.2cm} \forall k\in \mathbb{R} \label{eq18}
\end{align}
where the binary variable $Y_k = 1$, if all the PMs on $r_k$ are offline and hence the rack (along with cooling system and ToR switch) is going to shutdown in time-slot $t+1$. After all, in the following, we describe the formulation of the cost components in the proposed reliability-aware server consolidation model.\par

\subsection{Energy Cost ($C^{ene}$)} Let $\rho$ be the electricity charge measured in dollars per kWh. The total energy cost of the DC in time-slot $t+1$ is defined as $C^{ene}$ and is calculated using the following equation:
\begin{align}
    C^{ene} = \rho  \big( \sum_{i =1}^{|\mathbb{P}|} \nolimits c^{pm}_i + \sum_{j=1}^{|\mathbb{R}|} \nolimits c^{rack}_j + c^{mig} \big) \label{eq1}
\end{align}

\textit{PMs Energy Cost ($c^{pm}$):} In the above equation, $c^{pm}_i$ is the amount of energy consumption by the $i^{th}$ PM which is denoted by $p_i$. Now, the amount of this value has to be calculated. As it's known, CPU, memory, storage, and network are the main power consumers in a PM. However, studies show that CPU has the largest effect on a PM power consumption \cite{fan2007power, raghavendra2008no}. Also, these studies show that there is a linear power-to-frequency relationship for a PM. Moreover, it is shown in these studies that an idle PM consumes about 70\% of the power consumed by the PM running at the full CPU speed \cite{chen2008energy}. Therefore, to calculate the power consumption of a PM, we use the following power model \cite{esfandiarpoor2015structure}:
\begin{align}
    Power_i (\theta) = K  Power_{i}^{max} + (1-K)  Power_{i}^{max}  \theta \label{eq2}
\end{align}
\noindent where $Power_i$ denotes power consumption with respect to the CPU utilization $\theta$ of $p_i$. Notably, we consider the CPU utilization, as the PM utilization (i.e. $p_i \text{ } Utilization =  \sum_{j=1}^{|\mathbb{V}|} \nolimits S_{ji} R_j^{cpu} / \bar{C}^{cpu}_i$). As explained before, to show the relation of power consumption in idle and maximum utilization mode, we consider $K=0.7$ \cite{esfandiarpoor2015structure}. Also, energy consumption is the product of power consumption and time duration for power usage. Hence, the energy consumption of $p_i$ during time-slot $t+1$ is calculated as:
\begin{align}
    c^{pm}_i = \tau (1-(F_i^{00}+F_i^{10}))  Power_i (\theta) \label{eq3}
\end{align}
Notably, the value $c^{pm}_i$ must be 0 if $p_i$ is either remaining offline ($F_i^{00}=1)$, or will power off in $t+1$ ($F_i^{10}=1$). \par

\textit{Racks Energy Cost ($c^{rack}$):} The second term in Eq. (\ref{eq1}) is the energy consumption by the $j^{th}$ rack which is denoted by $c^{rack}_j$. In the proposed model, in addition to PMs, we try to turn off idle racks (as a result, the ToR switch and cooling device in a rack will turn off) to save more energy. Hence, the energy consumption of the $r_j$ is defined as:
\begin{align}
    c^{rack} = \sum_{k=1}^{|\mathbb{R}|} \nolimits Y_k (c^{ToR}+c^{cooling}) \label{eq4}
\end{align}
Notably, in this paper, both of ToR and cooling energy consumptions are considered as constant values.

\textit{VM Migration Cost ($c^{mig}$):} The last term in Eq. (\ref{eq1}) is $c^{mig}$, which is defined as the total energy consumption of VM migrations during server consolidation. VM migrations consume non-negligible energy, which increase linearly with the network traffic of migrating $v_i$ \cite{liu2013performance}. The memory size of the migrating VMs is the main part of this transferred data. Also, the distance between source and destination PM is worth to consider. For instance, migrating a VM between two PMs in a rack is more energy efficient than between east and west sides of the DC. Therefore, we define a VM migration cost matrix, $M_{\mathbb{P} \times \mathbb{P}}$ which each cell $M_{ij}$ denotes the cost of VM migration between $p_i$ and $p_j$. The value of these cells are a function of the memory of the VM that is determined to be migrated, and the distance between source and destination PMs. Therefore, the total VM migration cost can be written as follows:
\begin{align}
    c^{mig} = \sum_{i=1}^{|\mathbb{V}|} \nolimits \sum_{j=1}^{|\mathbb{P}|} \nolimits \sum_{k=1}^{|\mathbb{P}|} \nolimits S'_{ij} S_{ik} M_{jk} \label{eq5}
\end{align}

\subsection{Reliability Cost ($C^{rel}$)} We consider the reliability impacts of on-off cycles on wear-and-tear (disk and CPU) on PMs that are determined to be turned off, based on decrease of mean time to failure (MTTF) models in \cite{deng2014reliability}. Notably, we assume identical $MTTF$ for all DC PMs. We define $C^{rel}$ as the total reliability cost of the DC in time-slot $t+1$:
\begin{align}
    C^{rel} = \omega  \big( \sum_{i=1}^{|\mathbb{P}|} \nolimits F_i^{10} (c_i^{disk} + c_i^{cpu})  \big) \label{eq10}
\end{align}
where $c^{disk}_i$ and $c^{cpu}_i$ are the reliability degradation costs due to on-off cycles for PM disk and CPU, respectively. In fact, the disk and CPU reliability costs are applied for PMs that are turning off in time-slot $t+1$. Also, $\omega$ is the reliability utility per unit of time, which is defined as the ratio of the dollar cost to $MTTF$. For example, suppose the cost of a PM is 5,000 dollars and its average lifetime is 3 years. Then, $\omega$ would be $\frac{5,000}{3 8,760} = 0.1902$ (about 19 cents per hour). Now, let us define the reliability cost components in Eq. (\ref{eq10}).

\textit{Disk Reliability Cost ($c^{disk}_i$):} As it is reported in \cite{vishwanath2010characterizing}, 70\% of server failures are due to disk faults. Therefore, we first focus on describing the disk reliability cost. Start/stop cycles are recognized as the most important factor that cause reliability degradation in disks \cite{xie2008sacrificing}. So, the annual failure rate (AFR) with disk start/stop frequency $f$ is empirically formulated as:
\begin{align}
    AFR(f)= \delta e^{-5} f^2 - \varrho e^{-4}f + \varphi e^{-4}, \text{ } f \in [0,1600] \label{eq11}
\end{align}
where $\delta=1.51$, $\varrho=1.09$, and $\varphi=1.19$ \cite{xie2008sacrificing}. As AFR is the hours $H$ per year to the mean time between failures ($MTBF$), and $MTTF = MTBF $, by increasing AFR, the cost of (decreased $MTTF$) disk start/stop cycle for $p_i$ can be calculated as:
\begin{align}
    c^{disk}_i = \frac{H}{AFR_i(f)} - \frac{H}{AFR_i(f+1)} \label{eq12}
\end{align}

\textit{CPU Reliability Cost ($c^{cpu}_i$):} The damage accumulates with each CPU thermal cycle \cite{srinivasan2005lifetime}. Hence,  the increasing difference in temperature due to on-off thermal cycles decrease the CPU $MTTF$, which is this amount is proportional to $\big( \frac{1}{T_{avg}-T_{amb}}  \big)^q$, where $T_{avg}$ is the average CPU temperature, $T_{amb}$ is the ambient temperature (we assume $T_{amb}=298^{\circ}$ Kelvin), and $q$ is the constant Coffin-Manson exponent, suggested to be 2.35 \cite{srinivasan2005lifetime}. Thus, the CPU cost of turning off a PM $p_i$ in next time-slot ($t+1$) is calculated as:
\begin{align}
    c^{cpu}_i = \bigg[ \big( \frac{1}{T^i_{avg} - T_{amb}} \big)^q  \bigg].MTTF \label{eq13}
\end{align}
where $T^i_{avg}$ is the average CPU temperature after consolidation, and $MTTF$ is mean time to failure for a PM (e.g. 3 years).

\subsection{Reliability Gain ($G^{rel}$)} PMs lifetime can be conserved by turning the idle PMs off \cite{bodik2008case}. Therefore, the total reliability gain $G^{rel}$ for VM-to-PM mapping in time-slot $t+1$ is applied to two groups of PMs: First, the PMs that are turned off in time-slot $t$ and are determined to remain offline for time-slot $t+1$ (i.e. $F^{00}_i=1$). Second, the online PMs that are determined to be turned off in time-slot $t+1$ (i.e. $F^{10}_i=1$). Hence, $G^{rel}$ can be defined as the product of the total number of these PMs, $\omega$, and time-slot duration $\tau$:
\begin{align}
    G^{rel} = \omega  \tau  \sum_{i=1}^{|\mathbb{P}|} \nolimits (F_i^{00} + F_i^{10}) \label{eq14}
\end{align}
Putting it all together, the proposed optimization model for reliability-friendly server consolidation is presented as follows:
\begin{align}
    &\textbf{Minimize} \hspace{1cm}  \alpha \widehat{C^{ene}} + \beta \widehat{C^{rel}} - \gamma \widehat{G^{rel}} \label{eq19}  \\
    &\textbf{s.t.} \hspace{1cm} \nonumber \text{constraints } (\ref{eq6})-(\ref{eq1}), (\ref{eq3})-(\ref{eq10}),(\ref{eq14})\\
    &\textbf{vars.} \nonumber \hspace{.8cm}  S_{ij}, X_{jk}, Y_k , F_i^{10}, F_i^{00} \in \{ 0 , 1 \}, \nonumber \\
    & \hspace{1.45cm}\text{ and } c^{mig}, c^{pm}, c^{rack}, C^{rel}, G^{rel} \geq 0
\end{align}
where $\alpha$, $\beta$, and $\gamma$ are weighting factors (between 0 and 1). It is obvious that the proposed optimization model is in form of MILP which is NP-complete in general. On the other hand, in the above model, $\widehat{C^{ene}} = C^{ene} / C^{ene}_{ub}$, $\widehat{C^{rel}} = C^{rel}/C^{rel}_{ub}$, and $\widehat{G^{rel}} = G^{rel}/G^{rel}_{ub}$ are the normalized (0$\sim$1) values for cost components. However, to be able to normalize these cost components, we present three theorems to estimate the upper bound value for each of these cost components, and then, we use them to normalize the cost components values.

\begin{table*}[t]
\centering
\caption{The impact of different weighting factors on DC metrics}
\label{table2}
 \begin{tabular}{|c|c|c|c|c|c|c|c|c|} 
 \hline
 $\alpha$ & $\beta$ & $\gamma$ & \#Active Racks & \#Active PMs & \#VM Migrations & $C^{ene}$ & $C^{rel}$ & $G^{rel}$ \\
 \hline
 0.2 & 1.0 & 1.0 & 7 & 23 & 4 & 15224 & 2.5 & 0.72  \\ \hline
 1.0 & 0.2 & 1.0 & 4 & 13 & 23 & 9227 & 38.8 & 1.52\\ \hline
 1.0 & 1.0 & 0.2 & 4 & 16 & 19 & 9851 & 25.87 & 1.28\\
 \hline
 \end{tabular}
\end{table*}

\noindent \textbf{Theorem 1:} \textit{The upper bound value for $C^{ene}$ can be calculated as: $C^{ene}_{ub} = \rho  \tau  \big( |\mathbb{R}| c^{rack} + c^{pm}_{max} + |\mathbb{V}| c^{mig}_{max} \big)$.}
\begin{proof}
The energy consumption in time-slot $t+1$ is maximum when the DC has a number of specific conditions. Firstly, all the DC racks (set $\mathbb{R}$) should be active. Likewise, the number of online PMs should be maximum, which in this case, maximum energy consumption of PMs equals to:
\begin{align}
    c^{pm}_{max} &= \sum_{i=1}^{|\mathbb{P}| - \epsilon} \nolimits Power_i(\lfloor |\mathbb{V}|/|\mathbb{P}| \rfloor * R^{cpu} / \bar{C}^{cpu}) \nonumber \\
    & + \sum_{j=1}^{\epsilon} \nolimits Power_j(\lceil|\mathbb{V}|/|\mathbb{P}| \rceil * R^{cpu} / \bar{C}^{cpu}) \nonumber
\end{align}
where $\epsilon = |\mathbb{V}| - \lfloor |\mathbb{V}|/|\mathbb{P}| \rfloor * |\mathbb{P}|$ is the number of PMs that host $\lfloor |\mathbb{V}|/|\mathbb{P}|\rfloor + 1$ VMs. Additionally, VM migration costs should be maximum. To achieve this, in the worst case, maximum number of migrations should happen which equals to $|\mathbb{V}|$ and each with a cost equals to $c^{mig}_{max}$ (considering migration cost with $max(M_{|\mathbb{P}| \times |\mathbb{P}|})$). Hence, the upper bound for the value of DC energy consumption can be obtained using $C^{ene}_{ub} = \rho  \tau  \big( |\mathbb{R}| c^{rack} + c^{pm}_{max} + |\mathbb{V}| c^{mig}_{max} \big)$.
\end{proof}
\noindent \textbf{Theorem 2:} \textit{The upper bound value for $C^{rel}$ can be achieved using: $C^{rel}_{ub} = \sum_{i=1}^{(|\mathbb{P}| - \mathscr{P} )} \nolimits c^{disk}_i + c_i^{cpu}$ \textit{, where} $\mathscr{P} = \lceil \frac{\sum_{j=1}^{|\mathbb{V}|} \nolimits R^{cpu}_j}{\bar{C}^{cpu}} \rceil$.}
\begin{proof}
According to Eq. (\ref{eq10}), the maximum reliability cost achieve when CPU and disk reliability costs are applied to maximum number of PMs, i.e. maximum number of PMs should turn off, which means, all the VMs must be stored on minimum number of PMs. Considering the defined homogeneous DC, the lower bound for the number of PMs equals to $\mathscr{P} = \lceil \frac{\sum_{j=1}^{|\mathbb{V}|} \nolimits R^{cpu}_j}{\bar{C}^{cpu}} \rceil $ where $R^{cpu}_j$ is the VM $v_j$ utilization, and $\bar{C}^{cpu}$ is the total CPU capacity of identical PMs (in terms of cores, or MIPS). Thus, the upper bound value for reliability cost can be estimated as $ C^{rel}_{ub} = \sum_{i=1}^{(|\mathbb{P}| - \mathscr{P} )} \nolimits  c^{disk}_i + c_i^{cpu} $.
\end{proof}
\noindent \textbf{Theorem 3:} \textit{The upper bound value for reliability gain can be obtained using: $G^{rel}_{ub} = (|\mathbb{P}| - \mathscr{P})  \omega  \tau$ \textit{, where} $\mathscr{P} = \lceil \frac{\sum_{i=1}^{|\mathbb{V}|} \nolimits R^{cpu}_i}{\bar{C}^{cpu}} \rceil$.}
\begin{proof}
According to Eq. (\ref{eq14}), the reliability gain for duration $\tau$ is maximum, when the maximum number of PMs is turned off, i.e. all DC VMs must be hosted on minimum number of PMs. Therefore, we obtain the upper bound value for reliability gain in time-slot $t+1$ using $ G^{rel}_{ub} = (|\mathbb{P}| - \mathscr{P})  \omega  \tau $.
\end{proof}

\section{Performance Evaluation}
To simulate and assess the performance of the proposed mathematical model in Eq. (\ref{eq19}), we utilize MATLAB simulation software. We consider homogeneous PMs equipped with a processor with performance equivalent to 2,000 Million Instructions Per Second (MIPS), 10 GB of RAM, and 1 GB of network bandwidth, which their maximum power usage is $P^{max}=300$ $W$ \cite{esfandiarpoor2015structure}. Also, for simplification purposes, the we considered homogeneous VMs in the simulations which require a processor with performance of 500 MIPS and 612 MB of RAM, as Amazon EC2 Micro-Instance VM \cite{beloglazov2012optimal}. However, the problem can be easily extended to support multiple VM types.  Moreover, initially the VMs are allocated to random PMs according to the resource requirements of them. Additionally, suppose each rack of the DC hosts a number of PMs, and equipped with a rack-based cooling system with $c^{cooling}=950$ $W$ \cite{hpspec}, and a ToR switch with $c^{ToR}=366$ $W$ \cite{hpspec}. Finally, we consider the duration of a time-slot as $\tau = 0.5$ hour. All the experiments run on a computer running Microsoft Windows 10 Pro x64 with an Intel Core i7 Q740 processor and 10 GB of RAM. \par

\begin{figure}[t]
\centering
\includegraphics[width=.83\linewidth]{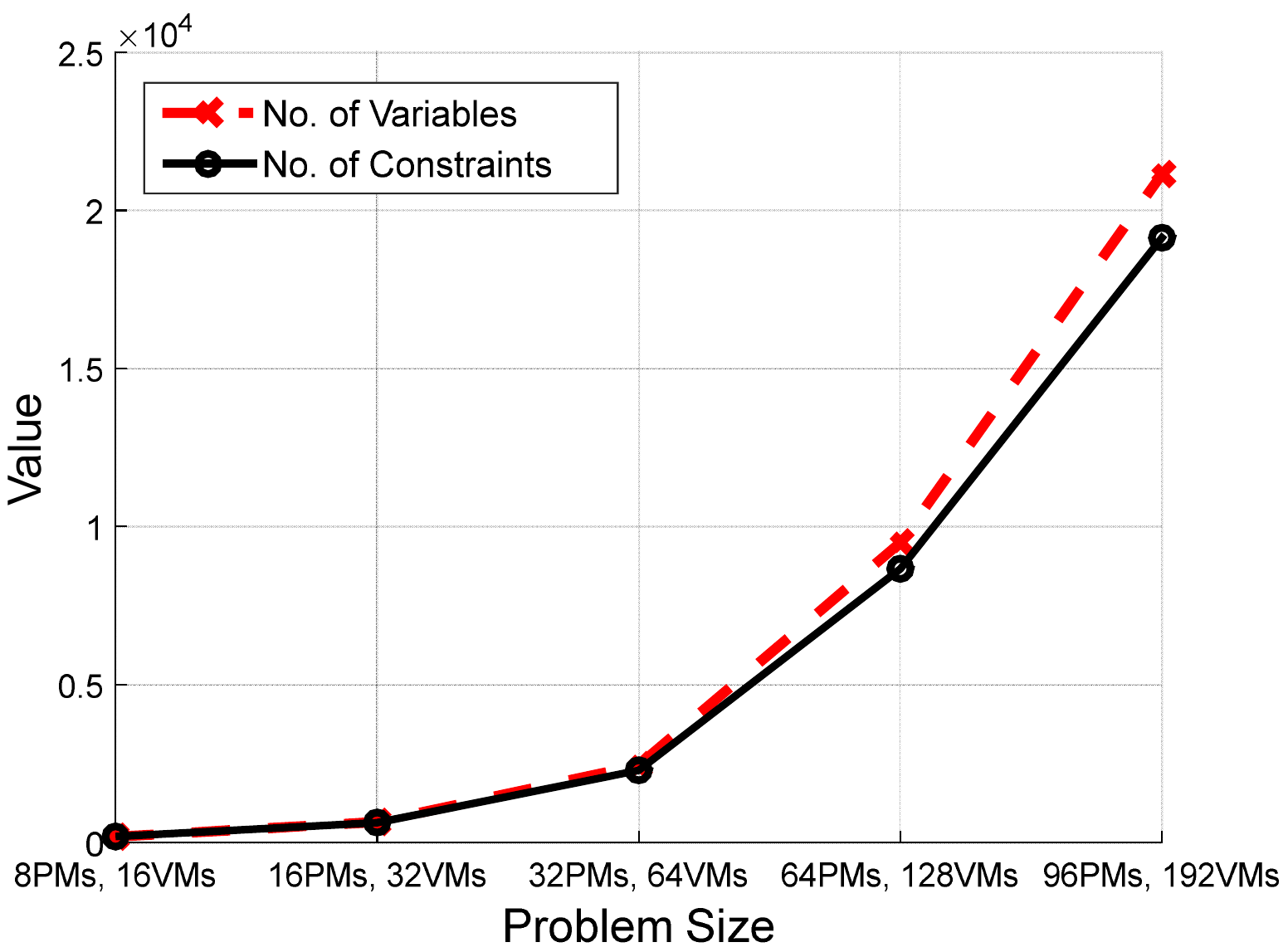}
\caption{Number of variables and constraints for different problem sizes}
\label{fig3}
\end{figure}
\begin{figure}[b]
\centering
\includegraphics[width=.83\linewidth]{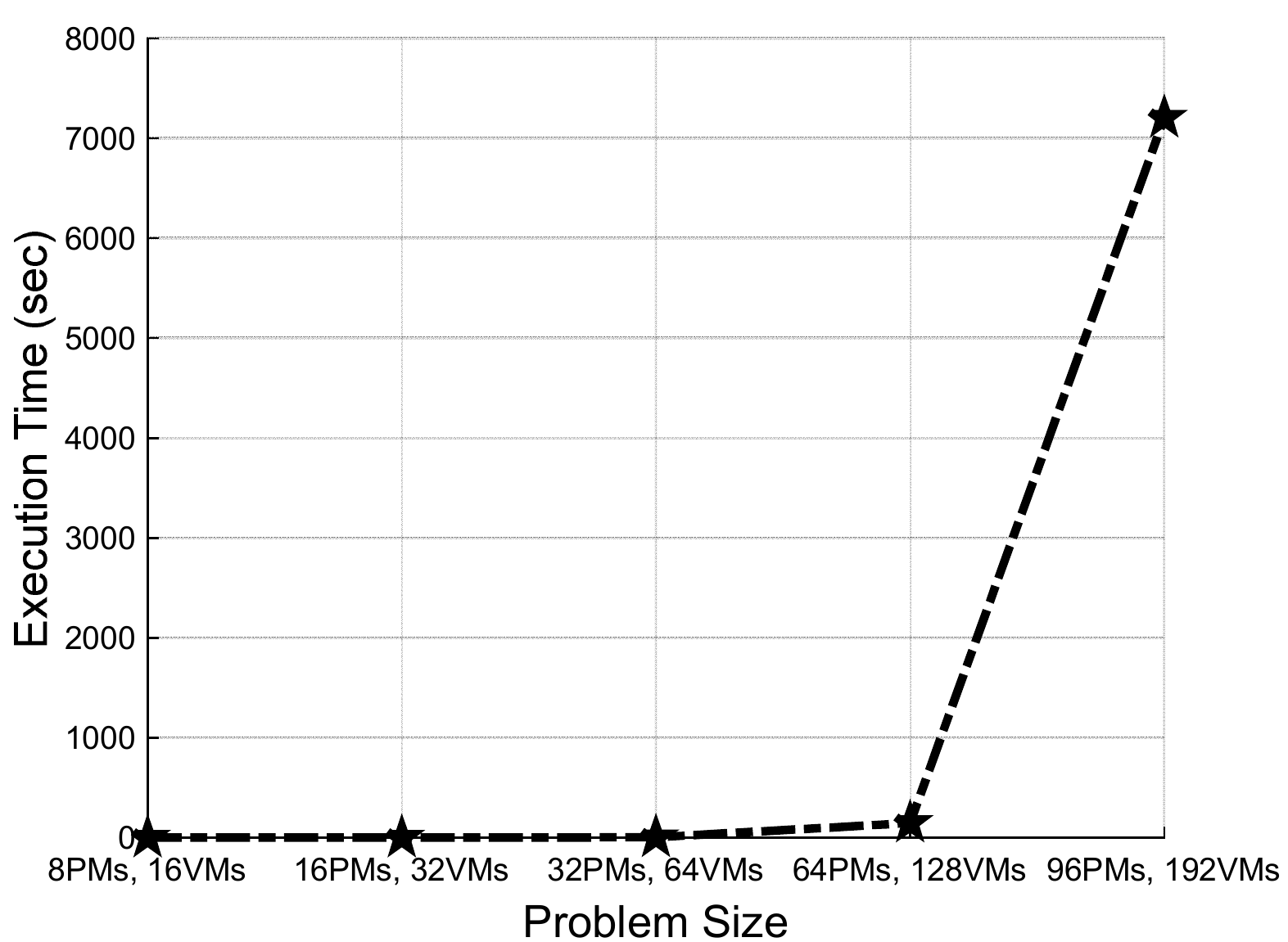}
\caption{Execution time for different problem sizes}
\label{fig4}
\end{figure}
We firstly run the model for different number of PMs with random initial VM placement to evaluate the number of constraints, variables, and the runtime of the proposed approach. As it is depicted in Fig. \ref{fig3}, the number of total variables and constraints are increasing exponentially with the growing size of PMs. Additionally, Fig. \ref{fig4} shows that the execution time is also growing exponentially with increasing the size of problem. Considering the real-time nature of the cloud DCs, developing a solution with high scalability and low overhead properties is mandatory. Therefore, according to observations in Fig. \ref{fig3} and \ref{fig4}, providing heuristics or meta-heuristics for solving this problem looks promising, which is the main objective of our future work. \par

In our next experiment, we consider an initial scenario with 32 PMs in 8 racks which host 52 VMs. To investigate the impact of different weighting factors (i.e. $\alpha$, $\beta$, and $\gamma$) on the problem parameters such as number of active PMs, number of VM migrations, etc., we run the simulation for different values of weighting factors. Notably, we run each scenario for 10 random initial VM-to-PM mappings and report the average value of the outputs (see Table \ref{table2}). It can be seen in Table \ref{table2} that by considering $\alpha=0.2$, in fact, the importance of energy cost is lower than the others. Thus, a lower number of VMs is migrated and as a result, the number of active racks and PMs are increased and hence, the DC consumes more energy. On the other hand, by setting $\beta=0.2$, we let the reliability cost to increase. Hence, more PMs are turned off and the energy cost decrease. Finally, it is demonstrated then by setting $\gamma=0.2$, the reliability gain and hence turning off the PMs have a small effect on the objective function. Therefore, a few more PMs are utilized and energy consumption slightly increase. \par

\begin{figure}[t]
\centering
\includegraphics[width=1\linewidth]{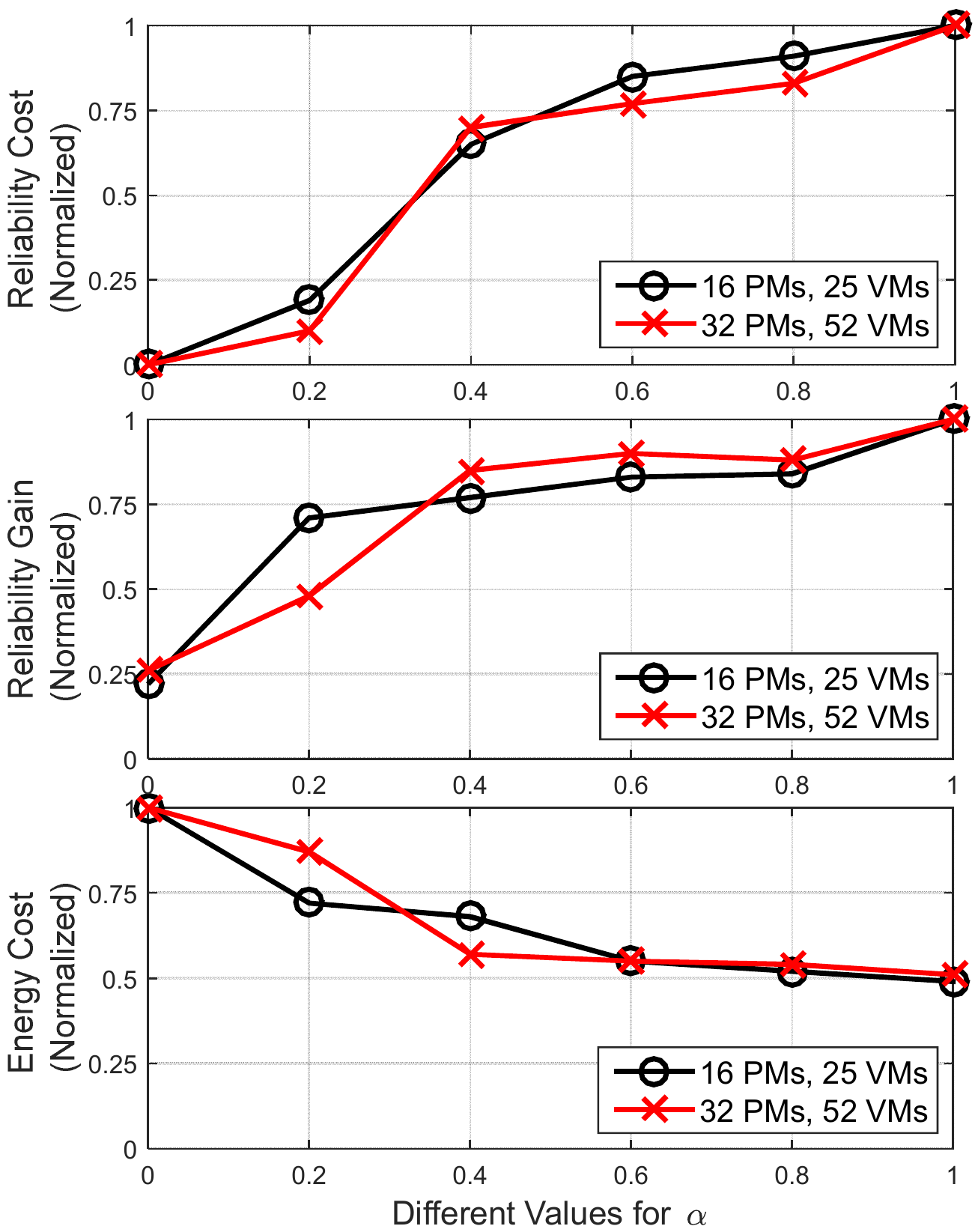}
\caption{Impact of different values for $\alpha$ on DC costs}
\label{fig5}
\end{figure}

As our final experiment, we consider two scenarios: 16PMs/25VMs and 32PMs/52VMs. We then run the simulation and increase the $\alpha$ values from 0 to 1 step-by-step. As it is illustrated in Fig. \ref{fig5}, the model has a similar behaviour for both scenarios. In fact, the relation between cost components are depicted clearly. When $\alpha$ increases, the energy consumption decreases gradually. Obviously, to achieve this energy consumption reduction, more and more PMs/Racks should be turned off. This will cause the reliability cost and also reliability gain to increase.

\section{Conclusion and Future Work}
Today, the energy consumption of Cloud datacenters (DCs) is one of the most important issues in technology world. Many techniques in different levels have been developed to make these DCs more energy-efficient, which one of them is server consolidation. In this technique, virtual machines (VMs) are packed on the minimum number of physical machines (PMs) and idle PMs are turned off to save energy. However, server consolidation could be utilized considering various parameters and factors, e.g. performance, network traffic, rack inlet temperature, and most recently, hardware reliability. Hardware reliability plays an important role in DC costs. Because firstly it could cause service outage which is expensive for DC managers. And secondly, it could highly affect the maintenance and replacement costs. In fact, in this paper, in addition to \textit{short-term} energy savings, we also took \textit{long-term} reliability and maintenance costs and lifetime of the PMs into account. In this work, we presented a reliability-aware server consolidation approach with the aim of minimizing total DC cost. This total cost consists of total DC energy including PMs, cooling and networking devices in each rack, and VM migration costs and also reliability costs including disk and processor on-off costs. Based on above considerations, we provided a mathematical model in form of Mixed Integer Linear Programming (MILP) which is NP-complete. We finally evaluated the performance of the proposed mathematical model using extensive numerical MATLAB simulations. \par

As future work directions, there are some interesting open challenges to discover. These days, Software Defined Networking (SDN) is an emerging paradigm which decouples network data plane and control plane. Using its centralized, network-wide abstraction of the control plane, SDN allows policies, configuration, and management of the DC to be applied in efficiently in short timescales. Therefore, in this area, there are some worthwhile problems to address, such as, developing SDN-based server consolidation and DC management frameworks, their SDN controller extensions, performance and resilience analysis. After all, considering the real-time nature of DC operation, providing heuristic/meta-heuristic approaches to find approximate solutions for the formulated problem can be an interesting challenge to explore.

\section{Acknowledgement}
We would like to thank anonymous reviewers for their valuable comments. This work was a part of Amir Varasteh's Masters thesis at Sharif University of Technology, Tehran, Iran. Also, the authors would like to thank Prof. Wolfgang Kellerer from the Chair of Communication Networks, Technical University of Munich, for his support.

\bibliographystyle{unsrt}
\bibliography{main.bib}

\end{document}